\documentstyle[twoside,fleqn,espcrc2,graphics]{article}

\begin{document} 
\newcommand{\cL}{{\mathcal{L}}}  
\newcommand{\ben}{\begin{equation}}
\newcommand{\een}{\end{equation}}
\newcommand{\bea}{\begin{eqnarray}}
\newcommand{\eea}{\end{eqnarray}}
\newcommand{\vev}[1]{\left\langle #1 \right\rangle}

\title{\begin{flushright}
\normalsize{SUSX-TH-98-018, hep-ph/9807372}
\end{flushright}
Constraints on Axion Models}

\author{Mark Hindmarsh\thanks{m.b.hindmarsh@sussex.ac.uk} 
and 
Photis Moulatsiotis\thanks{p.moulatsiotis@sussex.ac.uk}
  \address{Centre for Theoretical Physics \\
  University of Sussex \\ Brighton BN1 9QJ \\ U.K} }

\begin{abstract}
Two particular classes of axion models are presented, each one yielding a lower 
bound on the axion decay constant, based though on different considerations. In 
the first class only some, and not all, of the {\it right}-handed quarks have PQ 
charges, whereas in the second one the {\it left}-handed sector of the same 
quarks is taken into account as well. In the first case we find that bounds 
coming from astrophysics are significantly relaxed compared with those for the 
DFSZ. As for the second class, the astrophysical constraints proved to be less 
severe (with one exception), than those coming from FCNC processes.  
\end{abstract}
\maketitle

\section{INTRODUCTION}

The axion  \cite{PecQui77ab,WeiWil} is a particle which 
inevitably appears in one of the most appealing ways 
of solving the strong CP problem, which is the hypothesis 
that there is an extra global U(1) symmetry of the fermions 
of the Standard Model, known as the Peccei-Quinn (PQ) symmetry.  
The current benchmarks for realistic axion theories are 
the KSVZ \cite{KSVZ} and DFSZ \cite{DFSZ} models, to which most 
experiments are compared.  However, there are in principle 
many models, distinguished by different assignments 
of the quark and lepton transformation properties under the Peccei-Quinn 
symmetry.  This was realised, soon after the original 
axion model was proposed, by two groups \cite{PecWuYan86,KraWil86}.
However, all axion models based on identifying the electroweak scale with
the scale of PQ symmetry-breaking were quickly ruled out by accelerator
experiments and more seriously by astrophysical arguments based 
on energy loss from stars and supernovae 
(see \cite{AxRevs} for reviews).  The DFSZ and KSVZ models 
are examples of  ``invisible'' axion models, which break the
PQ symmetry with a singlet Higgs field.  The supervnova constraints 
\cite{SNConstr1,SNConstr2} bound the PQ scale to be above about 
$10^{10}$ GeV.  It took a few years for the models of Peccei, Wu and
Yanagida, and of Krauss and Wilczek, to be extended with a Higgs 
singlet \cite{GenNg89,CheGenNi95},  to be named ``variant'' axion models.

These models assign different PQ charges to different {\em right-handed} 
quarks: in the DFSZ model all quarks of a given chirality have the same charge.
Altering only the right-handed sector avoids any potential difficulties 
with flavour-changing neutral currents (FCNCs).  The question of motivation 
arises: why would we want to do this in the first place?  There are three 
reasons: firstly, models with only one PQ-charged right-handed quark do not 
have an axion domain wall problem \cite{Sik82}; secondly, one can use 
a PQ symmetry to distinguish the top quark and potentially explain its 
relatively large mass; and thirdly, one should explore all possibilities 
for axion models given the importance of the PQ solution of the strong
CP problem.

The cosmological domain wall problem arises in theories in which the 
Universe undergoes a phase transition at which the PQ symmetry is 
broken.  Axion strings form \cite{Sik82,AxStr}, which subsquently get connected 
by $2A$ domain walls, where $A$ is the QCD anomaly factor, which is either 
a half-integer or an integer.  Only in the case $A=1/2$ can the Universe 
avoid being subsequently dominated in energy density by the resulting 
network of walls and strings.  When  $A=1/2$ each string is connected to one
wall only, and can be drawn under the wall's surface tension to a neighbouring 
piece of anti-string, and annihilate.   We shall see that some of our 
models have $A=1/2$, and thus represent the simplest way so far of constructing 
viable models in the absence of inflation.

In the first part of this work, we shall investigate the supernova 
constraints on variant axion models, which are significantly different 
from those on the DFSZ model \cite{HinMou97}.  In the second part, in the spirit 
of 
inquiry, we consider the effect 
of different charges on the {\em left-handed} quarks. Here we must 
tackle full on the question of FCNCs, which constrain both the PQ 
scale, through the decays $K^+\to\pi^+ a$, and the masses of the other 
pseudoscalars in the theory, through the constraints on $B\bar B$ 
mixing.  In doing so we are greatly aided by the recent
work of Feng et al.\ \cite{Fen+98}, who comprehensively considered 
FCNC constraints on models with family symmetry. 
Our axion models are particular cases with an anomalous abelian
symmetry.

There are many types of model, depending on how the PQ charges are 
assigned \cite{HinMou98a,HinMou98b}.  In this work we examine a 
class of models with very strong constraints on the axion scale, 
stronger even than 
the supernova constraints with one exception, the case where only 
the top quark has a PQ charge.  The constraint on $B\bar B$ mixing requires 
that the other pseudoscalars should generically have masses greater 
than about $10^6$ GeV.  While not enough to rule out
the models, it certainly renders them unattractive, as one has to
prevent this mass scale from leaking into the Standard Model sector 
via radiative corrections.

\section{VARIANT AXION MODELS}
Variant axion models have two Higgs doublets $\phi_1$ and $\phi_2$ and one
Higgs singlet $\phi$, whose vacuum expectation values are
\bea
\vev{\phi_1} = \frac{1}{\sqrt 2}\left( \begin{array}{c} v_1 \\ 0 \end{array}
\right), 
\quad 
\vev{\phi_2} = \frac{1}{\sqrt 2}\left( \begin{array}{c} 0 \\ v_2 \end{array} 
\right), &&\nonumber\\
\vev{\phi} = \frac{v}{\sqrt 2}.&&\nonumber
\eea
In the DFSZ model, $\phi_1$ is used to give a mass to $u$-type quarks and 
$\phi_2$ to the $d$-type quarks, with both having the same PQ charge, which 
can be normalised to $\pm 1$, with the singlet having PQ charge 1/2.  In 
the variant models, only one of the Higgs fields has a PQ charge 
(which we can choose to be $\phi_1$), and this field is responsible for the
masses of the quarks with PQ charges.  The models are thus distinguished by 
which of the right-handed quarks is charged.  In Table \ref{ModelTable} 
we list the models (including three extra ones not considered in 
\cite{HinMou97}).
\begin{table}[hbt]
\setlength{\tabcolsep}{1.5pc}
\newlength{\digitwidth} \settowidth{\digitwidth}{\rm 0}
\catcode`?=\active \def?{\kern\digitwidth}
\caption{\label{ModelTable} The quarks with Peccei-Quinn charge 1 in the 
variant axion models considered in the text.  The rest of the quarks
(and leptons) have PQ charge zero.}
\begin{tabular}{ll}
\hline
Model & PQ-charged quark(s) \\
\hline
I & $u_R$ \\
II & $t_R$ \\
III & $u_R,t_R$\\
IV & $c_R$\\
V & $c_R,u_R$\\
VI & $c_R$, $t_R$\\
\hline
\end{tabular}
\end{table}
We cannot be any freer with the quark charge assigments without complicating 
the Higgs structure further.  Note that $\phi_2$ gives masses to both 
the $u$-type and $d$-type PQ-neutral fermions, just as in the Standard 
Model Higgs.

The QCD anomaly factor $A$ is given by 
\ben
A = \sum_q (Q_{Rq} - Q_{Lq}) t_q,
\een
where $q$ labels a chiral quark state in a representation with quadratic 
Casimir $t_q$.  For SU(3) triplets, $t_q=1/2$, and thus $A=1/2$ for 
models I, II and IV.  These are the models with no domain wall problem.

We define the axion scale $v_a$ through the kinetic term in the 
Lagrangian for the axion 
under the space-time dependent transformation $\exp(i\alpha Q)$, where 
$Q$ is the generator of the PQ charge
$
\cL_K = \frac{1}{2} v_a^2 (\partial \alpha)^2
$.  Note that many authors define an axion scale $f_a$, which is 
related to our $v_a$ by $f_a = v_a/2A$.  The advantage of this definition is
that the mass of the axion, and its coupling to gauge bosons, 
are inversely proportional to $f_a$.

The supernova constraint ultimately arises from nucleon bremsstrahlung 
in the nascent neutron star \cite{SNConstr1}:  if the axion couplings to
the nucleons is too strong, the resulting axion flux will change the 
picture of the supernova explosion, which is accurate enough to give 
reasonably good neutrino fluxes.  The limit \cite{SNConstr2} 
on the couplings $h_{an}$ and $h_{ap}$, the pseudoscalar couplings to 
the neutron and proton respectively, is quoted as
\ben
(2h_{an}^2+h_{ap}^2)^{1/2} < 2.85 \times 10^{-10},
\een
which is a fit to the boundary of a numerically determined excluded 
region.  There are many uncertainties in the calculation, and we have 
taken a conservative value from \cite{SNConstr2} which included 
possible many-body effects on the axion production rate.  Thermal 
pion conversion $\pi N \to N'a$ has been ignored, as it is thought to 
be unimportant.

\begin{figure}[ht]
  \centering
 \caption{\label{f:VaLims}
The lower bound on the Peccei-Quinn 
  symmetry-breaking scale $v_a$ for variant axion models described
  in the text and the DFSZ model, plotted as a function of $\beta$. 
  Here, $\tan \beta = v_2/v_1$, where $v_1$ and $v_2$ are the vacuum 
  expectation values of the two higgs fields giving masses to the
  fermions. The labels indicate which of the right-handed quarks have
  Peccei-Quinn charge.}
  \scalebox {0.4} {\includegraphics {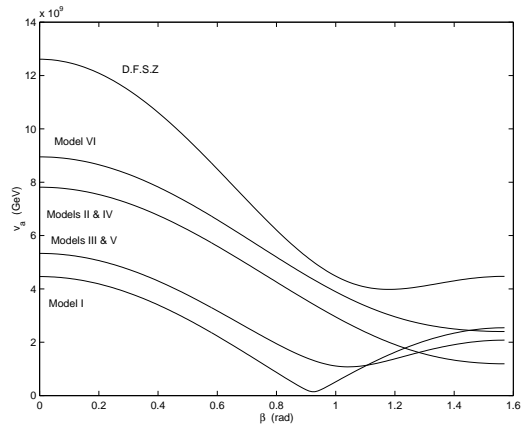}}
   \end{figure}

The calculation of the couplings  $h_{an}$ and $h_{ap}$ can be 
found in $\cite{HinMou97}$ for Models I to III, and can be straightforwardly 
extended to Models IV to VI.  We do not reproduce
them here, and instead merely exhibit the limit on 
$v_a$ which results from the relation $h_{aN} \sim m_N/v_a$ (Figure 
\ref{f:VaLims} ).  
The constraint on Model I, where the $u$ quark only is singled out, dips down to 
below $10^8$ GeV for a small range of values of the ratio $v_2/v_1$.  This 
is a viable axion string model, both in the sense that it has no domain wall
problem and that it is in no danger of violating even the most 
severe limits from the
energy density in axions radiated from strings,
which are dogged by theoretical uncertainty \cite{DavSheString,SikString}.  

\section{''DEVIANT'' AXION MODELS}
In this section we report of some forthcoming work \cite{HinMou98a} 
in which we attempt to 
give the left-handed quarks a variety of PQ charges.  A minimal departure from 
the DFSZ model is to have one or two quark doublets with PQ charge,
and the other uncharged, which  again 
defines six models. In the first 
three the `special' doublets are either the $(u{,}d)_L$, or $(c{,}s)_L$, or 
$(t{,}b)_L$, labeled by I, IV, II, and in the last three ones either 
$(u{,}d)_L$ and $(c{,}s)_L$, or 
$(u{,}d)_L$ and $(t{,}b)_L$, or $(c{,}s)_L$ and $(t{,}b)_L$ respectively, 
labeled by V, III, VI.  
We also have to assign PQ charges to the right-handed quarks. There are 
certain restrictions, as it turns out that if we try to introduce
too much variety there are too many zeros in the Yukawa coupling matrices, 
and we cannot reproduce the observed structure of the Cabibbo-Kobayashi-Maskawa 
matrix \cite{HinMou98b}.  One of the few realistic choices is to take 
the right-handed $u$-type quarks to have equal and opposite PQ charge to 
the left-handed quarks, and to give the right-handed $d$-type quarks 
PQ charge zero.

In order to do this we need at least three Higgs 
doublets   $\phi_n$ and one singlet 
$\phi$. If we allow four Higgs doublets, we have the possibility of making the 
model supersymmetric, as discussed below. 
The general structure of the Yukawa 
couplings is
\begin{equation}
\cL_Y=f^{n_u}_{ij}(\bar q^\prime_{Li}\phi_{n_u}u^\prime_{Rj})+f^{n_d}_{ij} (\bar 
q^\prime_{Li}\phi_{n_d}d^\prime_{Rj})+h.c
\end{equation}
where $n_u=1,3$, $n_d=2,4$ and $i,j=1,2,3$ are flavour indices.  We could of 
course use 
$\tilde\phi_3 = i\sigma_2\phi_3^*$ instead of $\phi_4$, but this would be 
forbidden in a supersymmetric model, as the superpotential (which determined 
the structure of the Yukawa terms) must be a function of $\phi_n$ and not
$\tilde{\phi}_n$. The most general transformation laws giving 
an abelian symmetry are 
\begin{eqnarray}
  u_{Ri}^\prime & \longrightarrow & e^{i\alpha {T}^u_{ij}}u_{Rj}^\prime 
\nonumber \\
  d_{Ri}^\prime & \longrightarrow & e^{i\alpha {T}^d_{ij}}d_{Rj}^\prime 
\nonumber \\
  q_{Li}^\prime & \longrightarrow & e^{i\alpha {T}_{ij}}q_{Lj}^\prime \\
  \phi_n & \longrightarrow & e^{iQ_n\alpha}\phi_n, \hspace{.5cm} n=1,2,3,4 .
\nonumber 
\end{eqnarray}
In all our models, $Q_1=Q_2=1$, and $Q_3=Q_4=0$.
Let us consider Model I,  
where the transformation matrices are given by
\begin{eqnarray}
{T}_{ij} &=& 
\left( \begin{array}{ccc}  
1 & 0 & 0 \\ 0 & 0 & 0 \\ 0 & 0 & 0
  \end{array} \right),  \nonumber\\
{T}^u_{ij} &=&\left( \begin{array}{ccc}  
-1 & 0 & 0 \\ 0 & 0 & 0 \\ 0 & 0 & 0
  \end{array} \right),  \nonumber\\
{T}^d_{ij} &=& 0.\nonumber 
  \label{e:T} 
\end{eqnarray}
The implementation of 
these symmetries fixes $f^n_{ij}$ to have zeros in certain entries. 
The only non-zero entry in $f^1_{ij}$ is $f^1_{11}$, and we must also have 
$f^3_{i1} = f^3_{1j} = 0$.  For the $d$-type Yukawa couplings, 
we need $f^2_{2j}= f^2_{3j} = 0$, and $f^4_{1j} = 0$. (Recall that 
it is $\phi_1$ and $\phi_2$ which have non-zero PQ charge).
More details can be found in \cite{HinMou98a,HinMou98b}.

Our model has tree-level FCNCs, and we must calculate the couplings 
in order to evaluate the constraints on the axion scale. In the flavour 
basis the relevant term of the QCD lagrangian is
\begin{eqnarray}
  \cL_{int}&=&-{\partial^\mu a^\prime\over 2v_a}[\bar u'_i\gamma_\mu 
((1-\gamma_5) 
T_{ij}\nonumber \\ &&+(1+\gamma_5)T^{u}_{ij} )u'_j \\
&&+\bar d'_i\gamma_\mu (1-\gamma_5) T_{ij}d'_j] \nonumber
\label{e:Lintfl}
\end{eqnarray}
The quark mass matrices are diagonalised by the transformations
\begin{eqnarray}
u'_{Ri}=U^R_{ij} u_{Rj}, \qquad d'_{Ri}=D^R_{ij} d_{Rj}, && \nonumber \\ 
u'_{Li}=U^L_{ij} u_{Lj}, \qquad d'_{Li}=D^L_{ij} d_{Lj}. && \nonumber
\end{eqnarray}
Applying these transformations to (\ref {e:Lintfl}) and going to the mass 
basis, the lagrangian takes the form
\begin{equation}
  \cL_{int}=\\
  {\partial^\mu a^\prime\over 2v_a}[2\bar u_i\gamma_\mu \gamma_5 
S^u_{ij} u_j-\bar d_i\gamma_\mu (1-\gamma_5) S^d_{ij} d_j]
  \label{e:Lintm}
\end{equation}
where $S^u={U^L}^\dag TU^L$ and $S^d={D^L}^\dag TD^L$. 
By definition, the Cabbibo-Kobayashi-Maskawa 
matrix is $V_{CKM}=U^{L\dag} D^L$, so it is obvious that
\begin{eqnarray}
S^d=V_{CKM}^\dag S^u V_{CKM}.
\label{e:TV}
\end{eqnarray}
Thus FCNCs are generally present, in both vector and axial-vector currents 
in the $d$-type quark sector.
Now, it is clear from the structure of the $u$-type Yukawa couplings that 
$U_L$ and $U_R$ have a block diagonal structure, with zeros in the 
first row and first column, and thus that $S^u = T$.
Hence one can easily work out $S^d$ from knowledge of the 
elements of the CKM matrix.

For the process $K^+\to \pi^+ a$ 
the interesting part of 
the interaction lagrangian (\ref{e:Lintm}) is the one giving the transition 
of $s\rightarrow d$ quarks. In this case 
\begin{equation}
  \cL_{int}=-{\partial^\mu a'\over 2v_a}
[\bar s\gamma_\mu (g_{sd}^V+g_{sd}^A\gamma_5) d+h.c.]
  \label{e:Lintsd}        
\end{equation}
where $g_{sd}^V$ and $g_{sd}^A$ are introduced, being 
the vector and axial vector parts of the $asd$ 
coupling. It is only the vectorial coupling which contributes to $K^+\to\pi^+ 
a$, 
which is 
\ben
g_{sd}^V=V_{ud}^* V_{us},
\een
where $|V_{ud}|\approx 0.98$ and $|V_{us}|\approx 0.22$.

The relevant experimental limit is 
${\mathrm{Br}}(K^+\rightarrow \pi^+ a)<3.0\times 
10^{-10} $ (at 90\% confidence) \cite{KpiaBound}. 
This leads to a lower bound on the axion energy 
scale \cite{HinMou98a}
\begin{equation}
v_a>1.7\times 10^{11}\times g_{ds}^V \hspace{1cm}\mathrm{GeV}.
\end{equation} 
Thus for Model I, $v_a>3.7\times 10^{10}$ GeV.  The limits for all models,
analagously derived, are listed in Table \ref{t:DevAxLims}, taken 
from \cite{HinMou98a}.
\begin{table}[hbt]
\setlength{\tabcolsep}{1.5pc}
\catcode`?=\active \def?{\kern\digitwidth}
\caption{\label{t:DevAxLims}Limits on the axion scale $v_a$ from 
the flavour-changing process $K^+\to \pi^+ a$.}
\begin{tabular}{lll}
\hline
Model & Charged & Limit\\
 & doublets &  (GeV)\\[1pt]
\hline
I & $ud$ & $3.7\cdot 10^{10}$\\
II & $tb$ & $6.1\cdot 10^7  $\\
III & $ud,tb$ & $3.7\cdot 10^{10}$\\
IV & $cs$ & $3.6\cdot 10^{10} $ \\
V & $cs,ud$ & $7.3\cdot 10^{10}$\\
VI & $cs,tb$ & $3.6\cdot 10^{10}$\\
\hline
\end{tabular}
\end{table}
With the exception of Model III, these bounds are all stronger than the 
corresponding supernova bound \cite{HinMou98a}.  The bound on Model III is
weak because of the smallness of $V_{td}$ and $V_{ts}$.

A theory with four Higgs doublets and a Higgs singlet has an additional three 
massive pseudoscalars. All of them can in principle mediate neutral meson 
mixing. $B\bar B$ is a particularly strong constraint, which forces 
the masses of the pseudoscalars to be greater than about $10^6$ GeV, 
\cite{HinMou98a} unless they are for some reason very weakly coupled to the $b$.  
Making a massive pseudoscalar is not a problem in principle, as there are in 
general terms such as $\lambda\phi^2\phi^T_1 \phi_2$, which contribute 
pieces of order $\lambda v^2$ to the masses. However, this leads 
to some kind of fine tuning in the Higgs potential, 
an ugly feature of these models which seems unavoidable.

One of the lessons we learn from these studies is that there is
plenty of freedom in assigning the Peccei-Quinn symmetry properties 
of the quarks, particularly in the right-handed sector.  If we wish 
to assign different charges in the left-handed sector, we must 
pay the price in inducing flavour-changing neutral currents mediated 
by the extra pseudoscalar bosons that are a result of the extra 
Higgs fields in such models.  To keep these at an acceptable level 
we must tune the Higgs potential such that these pseudoscalars have
masses greater than about $10^6$ GeV, while keeping the electroweak 
Higgs vacuum expectation value at $246$ GeV.


M.H.\ is supported by PPARC Advanced Fellowship  
B/93/AF/1642 and by PPARC grants  GR/L56305 and GR/L55759.

\end{document}